\documentclass[%
 reprint,
 amsmath,amssymb,
 aps,
]{revtex4-2}
\usepackage[utf8]{inputenc}
\usepackage{dcolumn}

\usepackage{bm}
\usepackage{mathptmx}
\usepackage{amsmath}
\usepackage{etoolbox}
\usepackage{multirow}
\usepackage{blindtext}
\usepackage{graphicx}
\usepackage{subcaption}
\usepackage[export]{adjustbox}
\usepackage{wrapfig}
\usepackage{comment}

\usepackage{array}
\usepackage[table]{xcolor}
\newcolumntype{s}{>{\columncolor[HTML]{AAACED}} p{3cm}}

\graphicspath{{images/}}
\usepackage{siunitx}

\usepackage{graphicx}
\usepackage{dcolumn}
\usepackage{bm}

\begin{document}

\preprint{APS/123-QED}

\title{Spin Phonon Coupling and Relaxation time in Lu(II) compound with 9.2GHz clock transition}

\author{Xiaoliang Zhang}
 \email{xiaolian.zhang@ufl.edu}
\author{Haechan Park}%

\affiliation{ 
Department of Physics and the Quantum Theory Project\\
University of Florida, Gainesville, FL 32611
}%




\date{\today}






\begin{abstract}

  Electron spin qubits operating at atomic clock transitions exhibit exceptionally long coherence times, making them promising candidates for scalable quantum information applications. In solid-state systems, interactions between qubits and lattice phonons are known to play a critical role in spin relaxation ($T_1$) and decoherence ($T_2$). In this work, we perform first-principles calculations on a Lu(II) complex spin qubit featuring a prominent clock transition. By employing advanced electronic structure methods, we quantitatively evaluate the influence of phonons on the hyperfine interaction, which serves as the primary spin–lattice coupling mechanism. Treating these phonon-induced variations as first-order perturbations, we apply the Redfield master equation to compute both $T_1$ and $T_2$, along with their temperature dependencies.
For $T_1$, we adopt a second quantization formalism to describe phonon interactions, while $T_2$ is evaluated by explicitly integrating acoustic phonon contributions across the full Brillouin zone. Our results reproduce the experimentally observed magnetic field dependence of $T_2$, including the coherence peak near 0.43 T, though the absolute values of $T_1$ and $T_2$ differ by one to two orders of magnitude. Analysis reveals that $T_1$ is primarily governed by longitudinal phonons, whereas $T_2$ is most strongly influenced by mid-wavelength, mid-energy acoustic modes. These findings provide a quantitative demonstration of the clock transition’s protective effect on spin qubit coherence and offer a transferable computational framework for evaluating spin-phonon interactions in other molecular spin qubits.
  
  
\end{abstract}

 \maketitle
\section{Introduction}

Electron spin relaxation and decoherence involve interactions among electron spins, nuclear spins, magnetic fields, and environmental factors. These processes are crucial for quantum systems using electron spins as qubits, affecting their coherence and stability. As significant progress has been made in mitigating decoherence from nuclear spin baths \cite{PhysRevB.65.205309,Saykin2002,PhysRevB.74.195301,PhysRevB.85.115303,PhysRevLett.110.160402}, electron spin decoherence arising from thermal motion of ions, usually referred to as
spin-phonon coupling, has become increasingly important.


Spin-phonon interactions have been known to cause spin relaxation and decoherence for almost a century. Various spin relaxation theories(\cite{doi:10.1021/jacs.1c05068,Murali2003APF}) have been applied to nuclear spin decoherence and electronic spin decoherence, demonstrating significant success.  In recent decades, spin-phonon interactions via various mechanisms, such as spin-orbit coupling, hyperfine interaction, and spin-nuclear coupling, significantly contribute to spin flips and decoherence in quantum dots and other solids\cite{STAVROU2021114605,PhysRevB.101.165302,PhysRevA.100.042309,PhysRevB.98.075403}.  These interactions have been theoretically explored, providing qualitatively insights into their roles and effects\cite{PhysRevLett.93.016601,Gomez-Coca2014}.

In recent years, first-principles calculations have been extensively applied to molecular and crystalline systems to quantitatively determine electronic densities and investigate electron spin relaxation. Notably, \textit{ab initio} approaches have enabled the study of phonon-induced spin relaxation in the presence of spin-orbit coupling, revealing the underlying roles of the Elliott-Yafet and Dyakonov-Perel mechanisms in spin decoherence within a many-body framework\cite{PhysRevB.101.045202,PhysRevLett.129.197201,PhysRevB.106.174404}.   Other studies have employed first-principles methods within the second quantization formalism to quantify the influence of phonon-induced modifications in Zeeman, hyperfine, and electronic spin dipolar interactions on direct spin relaxation processes~\cite{Lunghi2017,C7SC02832F,Escalera-Moreno2017}. These \textit{ab initio} methods have demonstrated strong predictive capabilities, yielding longitudinal spin-lifetime results ($T_1$) within an order of magnitude of experimental data~\cite{doi:10.1126/sciadv.aax7163}, thereby capturing the essential physics underlying spin relaxation.

Despite their success, previous first-principles approaches to spin-phonon interactions have primarily relied on quantum electrodynamics and second quantization formalisms. While these frameworks are generally adequate for estimating the longitudinal relaxation time ($T_1$), they often fail to quantitatively capture zero-energy transfer fluctuations that are essential for understanding spin decoherence and the transverse relaxation time ($T_2$) \cite{PhysRevB.101.045202,PhysRevLett.129.197201,PhysRevB.106.174404,Lunghi2017,C7SC02832F,Escalera-Moreno2017,doi:10.1126/sciadv.aax7163,Briganti2021,doi:10.1126/sciadv.abn7880}. In this work, we go beyond the conventional treatment by performing density functional theory (DFT) calculations not only for $T_1$, but also for $T_2$, through evaluating hyperfine interaction fluctuations induced by lattice vibrations.


Among the various types of spin qubits, molecular spin qubits based on atomic clock transitions have demonstrated exceptionally long coherence times in several experimental studies~\cite{Shiddiq2016,Kundu2022-kg,Gakiya-Teruya2025}. Atomic clock transitions occur at specific magnetic field regimes where quantum spin dynamics are intrinsically protected from environmental perturbations. This intrinsic robustness makes clock-transition-based spin qubits highly promising for scalable quantum architectures involving large arrays of spin qubits. Of particular interest is the Lu(II) complex  \((OAr^{\ast})_3\text{Lu}\), which hosts a spin qubit exhibiting remarkably long coherence times due to such a transition~\cite{Kundu2022-kg}. These characteristics make it a strong candidate for quantum information applications. Consequently, understanding how phonons influence both the longitudinal relaxation time ($T_1$) and the transverse decoherence time ($T_2$) in these systems is crucial for the development of robust, spin-based qubits.
 
In this work, we perform \textit{ab initio} calculations to obtain accurate electronic densities and their phonon-induced variations in the Lu(II) complex  \((OAr^{\ast})_3\text{Lu}\). Based on the one-phonon spin-phonon interaction and using the Redfield master equation, we compute the spin-lattice relaxation time $T_1$ under second-quantization formalism\cite{doi:10.1126/sciadv.aax7163}. 

\begin{figure}
    \centering
    \includegraphics[scale=0.24]{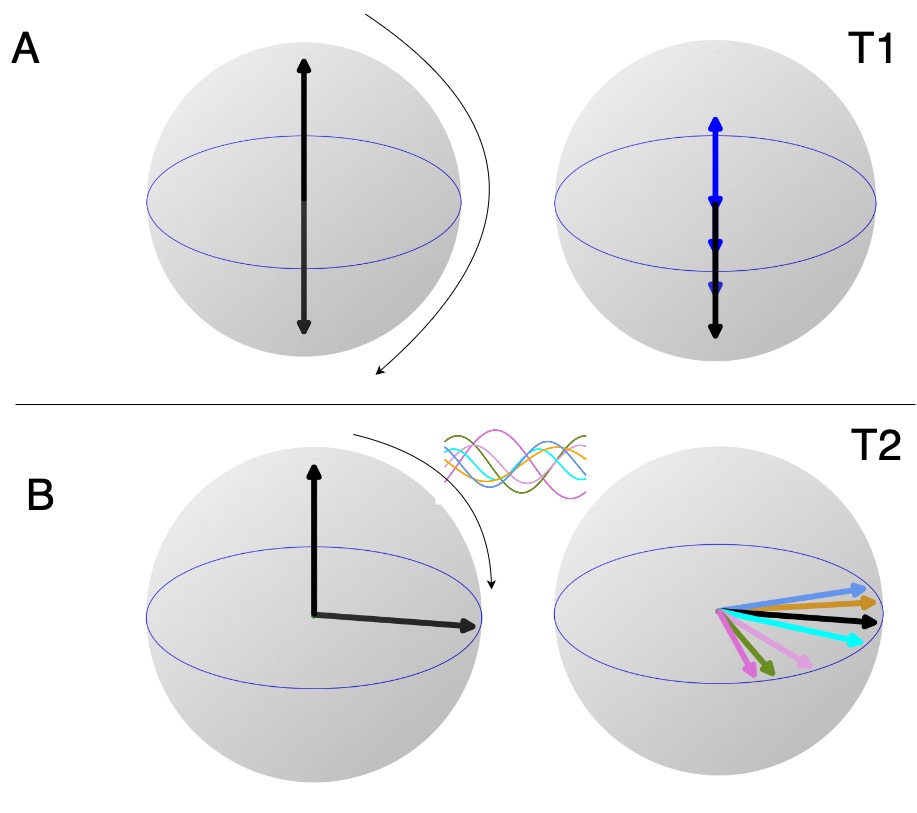}
    \caption{\textbf{Schematic illustration of the distinct relaxation mechanisms for \( T_1 \) and \( T_2 \).} 
    \textbf{A}, The longitudinal relaxation time \( T_1 \) characterizes the return of the spin from an excited state (e.g., spin-down) back to its equilibrium position (spin-up) along the \( z \)-axis after an external pulse. 
    \textbf{B}, The transverse relaxation time \( T_2 \) describes the dephasing of spins in the transverse plane following a \(\pi/2\) pulse. Due to interactions with a variety of phonon modes(as illustrated in the central inset) with different frequencies and phases, spins precess at slightly different rates, leading to the loss of phase coherence over time.}

    \label{schet1t2}
\end{figure}

$T_2$ consists of two components: the first arises from relaxation processes, while the second originates from dephasing due to fluctuations in the thermal environment\cite{PhysRevB.73.155311,Kornich2018phononassisted}, as the Fig.~\ref{schet1t2} demonstrates schematically. In this work, we employ a numerical approach to model fluctuations in hyperfine interactions induced by lattice vibrations, which are found to dominate the $T_2$ decoherence process. By integrating over all acoustic phonon branches throughout the entire Brillouin zone~\cite{YU2022111000}, we capture the intrinsic transverse spin dispersion caused by longitudinal hyperfine field fluctuations—identified as the primary mechanism driving spin decoherence in this system.


Altogether, our results underscore the significant role of phonon-induced decoherence, even in highly coherent spin systems, and demonstrate that both the trends and magnitude orders of $T_1$ and $T_2$ can be reliably predicted using first-principles methods. These findings deepen our understanding of spin dynamics in clock-transition systems and provide valuable guidance for the design of more robust and scalable qubit platforms.

\section{Clock transition and Spin-Phonon coupling}
The tris(aryloxide) Lu complex, $(OAr^{\ast})_3$Lu, facilitates the formation of a $4f^{14}5d^{1}$ Lu(II) complex. This complex exhibits a significant, nearly isotropic hyperfine interaction of approximately 3500 MHz, which enables a stable clock transition around 9.2 GHz. This clock transition contributes to an exceptionally long spin lifetime. Experimental measurements show a longitudinal relaxation time ($T_1$) of 1.6 ms and a transverse decoherence time ($T_2$) of about 12 $\mu$s at a temperature of 5 K\cite{Kundu2022-kg}. The corresponding spin Hamiltonian is presented below.

\begin{equation}\label{sph}
    \hat{H}=\mu_B B_0\cdot g_e\cdot \hat{S}-\mu_N g_n B_0\cdot \hat{I}+\hat{S}\cdot \overleftrightarrow{A} \cdot \hat{I} +\hat{I}\cdot Q \cdot \hat{I}
\end{equation}
\begin{figure}
    \centering
    \includegraphics[scale=0.15]{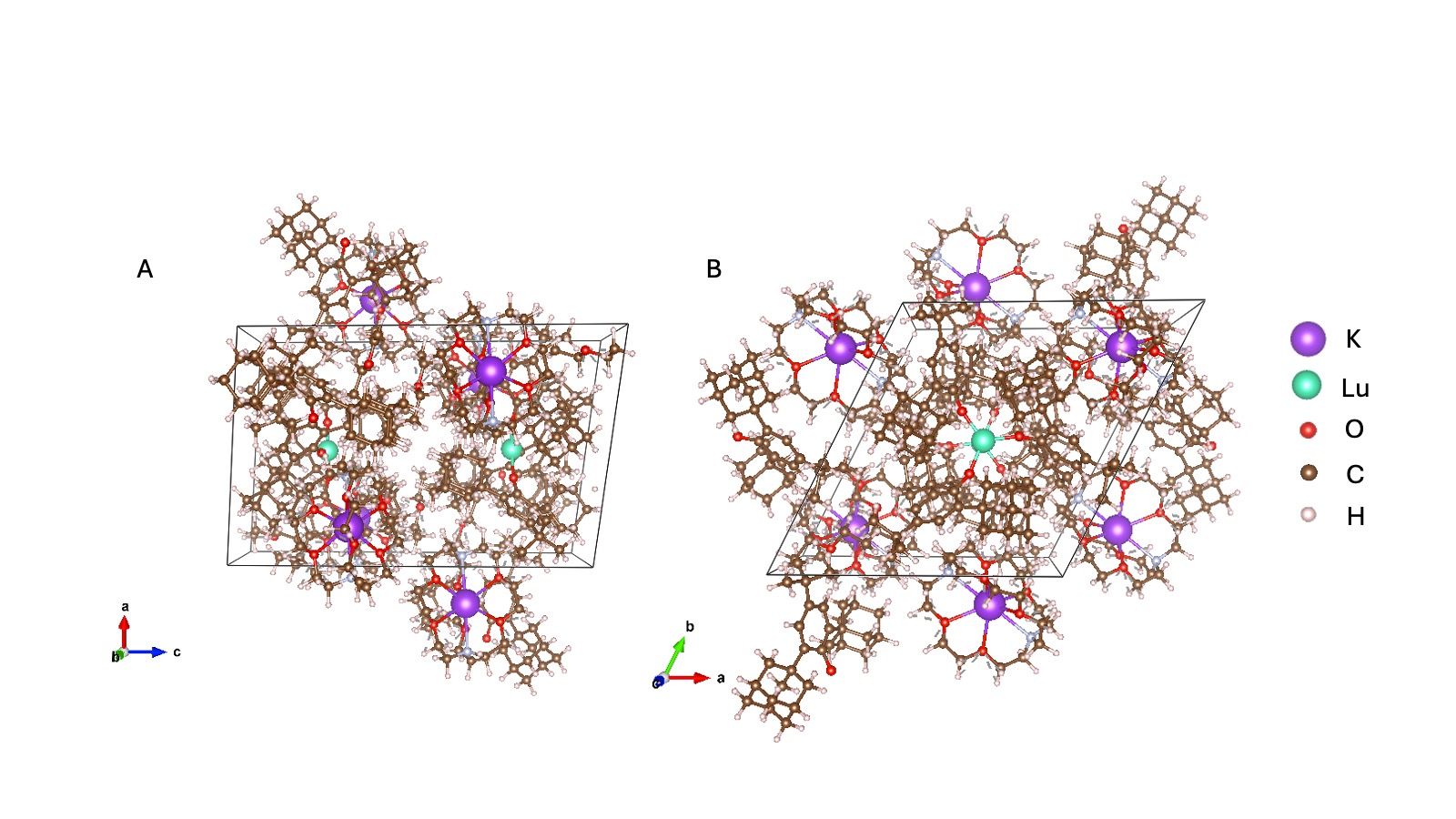}
    \caption{Crystal structure of Tris(aryloxide) Lu complex $(\mathrm{OAr}^{\ast})_3\mathrm{Lu}$. Lu atoms are shown in green. (A) View along the lattice vector $\mathbf{b}$. (B) View along the lattice vector $\mathbf{c}$. 
      }
    \label{lucom}
\end{figure}

\begin{figure}
    \centering
    \includegraphics[scale=0.24]{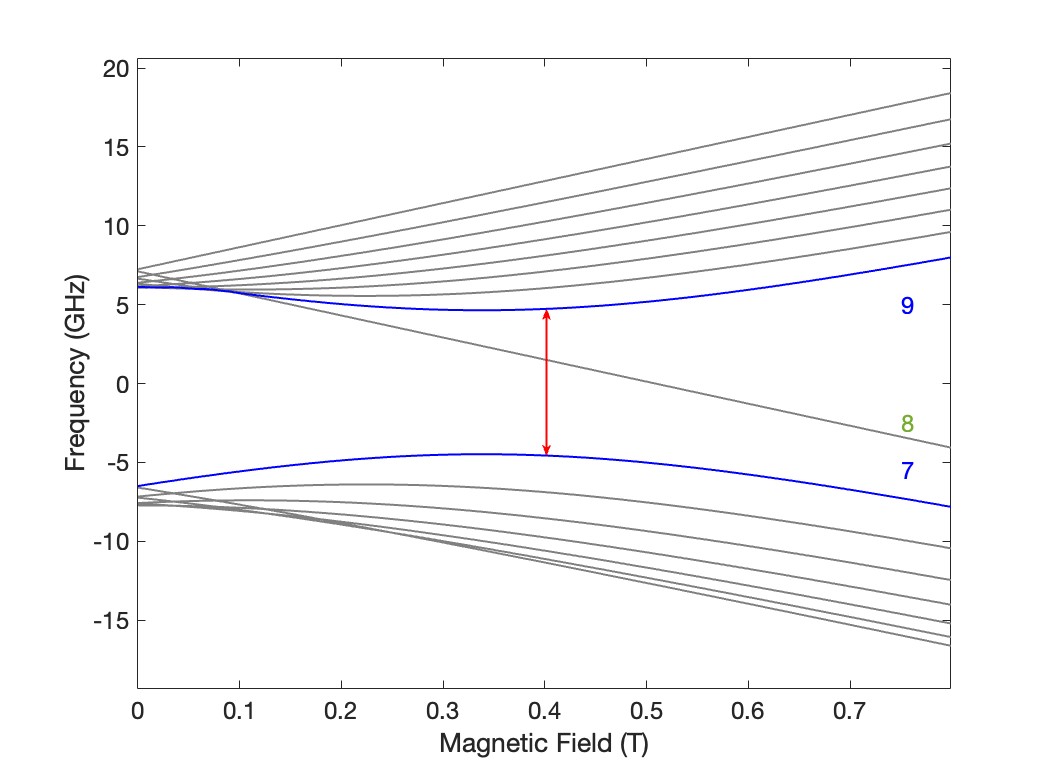}
    \caption{Energy level diagram of the 16 eigenstates arising from hyperfine coupling between the electron spin ($S = \frac{1}{2}$) and the nuclear spin ($I = \frac{7}{2}$) of the Lu(II) ion, plotted as a function of the external magnetic field. The clock transition occurs between states 7 and 9, highlighted by the two blue curves. The red arrow indicates the position of the clock transition at a static magnetic field of approximately 0.4 T.
      }
    \label{clocklu}
\end{figure}

The first and second terms in Eq.~\ref{lucom} represent the Zeeman interactions of the electron spin and the nuclear spin, respectively. The third term describes the hyperfine interaction, where the tensor $\overleftrightarrow{A}$ characterizes the coupling between the electronic spin $\mathbf{S}$ and the nuclear spin $\mathbf{I}$. In this molecule, $\overleftrightarrow{A}$ is diagonal. The final term accounts for the quadrupole interaction of the Lu nucleus. For the Lu atom, the nuclear spin is $I = \frac{7}{2}$, yielding $2I + 1 = 8$ nuclear spin states. When combined with a single unpaired electron in an $S$-orbital ($S = \frac{1}{2}$), the total spin Hilbert space consists of $8 \times 2 = 16$ states, governed by the spin Hamiltonian in Eq.~\ref{sph}. The clock transition occurs between state 7 and state 9, highlighted by the two blue curves in Fig.~\ref{clocklu}.

The eigenvalues and eigenstates of this Hamiltonian were computed using the open-source MATLAB package EasySpin~\cite{STOLL200642}. The parameters used in the simulation are summarized in Table~\ref{hamilpara}. The hyperfine tensor and Landé \(g\)-tensor were taken from experimental measurements~\cite{Kundu2022-kg}, while the nuclear quadrupole interaction parameter \(Q_{zz}\) was obtained from an independent study~\cite{89596ae1770f47d59ee96c48b2b10c8f}.

The resulting energy level diagram, derived from diagonalizing the spin Hamiltonian as a function of the static magnetic field, is shown in Fig.~\ref{clocklu}.

\begin{table}[h!]
 \caption{\textbf{Spin Hamiltonian Parameters.} 
$A_{xx}$, $A_{yy}$, and $A_{zz}$ are the diagonal components of the hyperfine tensor; $g_{xx}$, $g_{yy}$, and $g_{zz}$ are the diagonal components of the Landé $g$-tensor. The $A$ and $g$ tensor values are taken from experimental measurements~\cite{Kundu2022-kg}. $Q_{zz}$ denotes the quadrupole component of the nuclear spin~\cite{89596ae1770f47d59ee96c48b2b10c8f}.}
\label{hamilpara}
 \centering
 \begin{tabular}{|c|c|}
 \hline
   $A_{xx} $   & 3500 $\pm $ 50 MHz \\
   \hline
   $A_{yy} $   & 3500 $\pm $ 50 MHz\\
   \hline
   $A_{zz} $   & 3400 $\pm $ 50 MHz \\
   \hline
    $g_{xx} $   & 1.915 $\pm$ 0.002\\
   \hline
   $g_{yy} $   & 1.915 $\pm$ 0.002\\
   \hline
   $g_{zz} $   & 2.000 $\pm$ 0.002\\
    \hline
   $Q_{zz} $   & 100 $\pm $ 20 MHz\\
   \hline        
 \end{tabular}

 \end{table}

The hyperfine interaction comprises two primary components: the magnetic dipolar interaction and the Fermi contact interaction. The isotropic Fermi contact term arises from the direct interaction between the nuclear magnetic dipole moment and the electron spin density at the nucleus. It is non-zero only when there is a finite electron spin density at the nuclear position, typically associated with 
s
s-orbital electrons. In this material, the Fermi contact interaction dominates the hyperfine coupling due to the presence of unpaired electrons in the 
s
s-subshells~\cite{Kundu2022-kg}. As shown in Eq.~(\ref{fermi}), the strength of the Fermi contact interaction is directly proportional to the electron spin density at the nucleus.

\begin{equation}\label{fermi}
A_{iso,Fermi}=\frac{2\mu_{0}}{3h}\mu_N g_n\mu_B g_e\rho_s(\bf{R}_n)
\end{equation}

Here, \(\rho_s({\bf R}_n)\) represents the electron spin density at the nuclear position \({\bf R}_n\). Variations in \(\rho_s({\bf R}_n)\) due to lattice vibrations are the primary source of spin-phonon coupling in this system. These fluctuations induce a time-dependent perturbation \(A_1(t)\) in the spin Hamiltonian [see Eq.~(\ref{sph})], which drives both spin relaxation and decoherence processes.

We introduce spin-phonon coupling as the first order perturbative term in reciprocal space, which takes the form as follows:
\begin{equation}
    H_{sph}=\sum_{\alpha\Vec{q}}\frac{\partial A}{\partial Q_{\alpha\Vec{q}}} \hat{Q}_{\alpha\Vec{q}}=\sum_{\alpha\Vec{q}}A^{\alpha\Vec{q}}_1 \hat{Q}_{\alpha\Vec{q}}
\end{equation}


The quantity $A^{\alpha\vec{q}}_1$ represents the variation in the hyperfine tensor resulting from the collective atomic displacements associated with a phonon mode of momentum $\vec{q}$ in branch $\alpha$. It is defined as the derivative of the hyperfine tensor with respect to the collective phonon displacements $Q_{\alpha\vec{q}}$, and can be written as a linear combination of atomic displacements within the unit cell, as expressed in Eq.~(\ref{phodis}).

\begin{equation}\label{phodis}
    A^{\alpha\vec{q}}_1 = \frac{\partial A}{\partial Q_{\alpha\vec{q}}} = \sum_{i=1}^{N} \sum_{s=1}^{3} \sqrt{\frac{\hbar}{N_{\vec{q}} \omega_{\alpha\vec{q}} m_i}} \, L_{is}^{\alpha\vec{q}} \, \frac{\partial A_s}{\partial X_{is}}
\end{equation}

Here, $i$ indexes the atoms in the unit cell, with $N$ being the total number of atoms. The index $s$ runs over the three Cartesian directions ($a$, $b$, and $c$) within the unit cell. $X_{is}$ denotes the displacement of atom $i$ along direction $s$, and $L_{is}^{\alpha\vec{q}}$ is the phonon eigenvector corresponding to atom $i$ and direction $s$ for mode $(\alpha, \vec{q})$, obtained from the diagonalization of the dynamical matrix using the Phonopy package~\cite{phonopy-phono3py-JPCM}. $\omega_{\alpha\vec{q}}$ is the phonon frequency for the specified mode, $m_i$ is the atomic mass of atom $i$, and $N_{\vec{q}}$ is the number of $\vec{q}$-points sampled in the Brillouin zone.

The hyperfine tensor and its derivatives with respect to atomic displacements, $\frac{\partial A_s}{\partial X_{is}}$, were computed using the Vienna Ab initio Simulation Package (VASP)~\cite{PhysRevB.49.14251}. The investigated Lu(II) complex has the chemical formula C${120}$H${189}$KLuN$2$O${12}$ and contains two identical molecules per unit cell, resulting in a total of 650 atoms.

To evaluate Eq.~(\ref{phodis}), each atom in the unit cell was incrementally displaced along the three crystallographic directions ($a$, $b$, and $c$). Displacements ranged linearly from $-0.008$~Å to $0.008$~Å with 8 discrete steps per direction, amounting to $650 \times 3 \times 8 = 15{,}600$ separate DFT calculations. For each displacement step, density functional theory (DFT) calculations were carried out to determine the corresponding variation in the hyperfine tensor. The resulting discrete data were subsequently fitted to a second-order polynomial for each atomic site, enabling smooth interpolation of the hyperfine tensor as a function of atomic displacement.

The DFT calculations were performed using the PAW-PBE pseudopotentials within the generalized gradient approximation (GGA)\cite{PhysRevB.50.17953}. The plane-wave energy cutoff (ENCUT) was set to 600~eV, and the electronic self-consistent convergence threshold (EDIFF) was set to $10^{-8}$eV. Collinear spin polarization (ISPIN = 2) was enabled to account for the open-shell nature of the Lu(II) ion. Given that lutetium is a rare-earth element with localized $f$-electrons, the DFT+$U$ method\cite{PhysRevB.44.943,Rohrbach_2003} was employed with an effective Hubbard $U$ parameter of 4.7~eV, consistent with prior literature\cite{PhysRevB.94.014104}. 
The hyperfine tensors calculated from these \textit{ab initio} simulations show good agreement with experimental values, with a typical deviation within 30~MHz, confirming the reliability of the computational setup.



\section{Redfield theory}

The Redfield master equation of spin reduced density matrix\cite{Murali2003APF}
\begin{equation}\label{densm}
    \frac{d\rho^s_{ab}(t)}{dt}=\sum_{cd}e^{i(\omega_{ac}+\omega_{db})t}R_{ab,cd}\rho^s_{cd}(t)
\end{equation}
In which $a,b,c,d$ are the eigenstates of unperturbed Hamiltonian, and $\omega_{ab}$ stands for the energy difference between $a$ and $b$ state. Typically, it requires $\omega_{ac}+\omega_{db}=0$, which is so call secular approximation, corresponding to the assumption that relaxation time is much longer than the periods of any $\omega_{ac}$, then any $\omega_{ac}+\omega_{db}\not=0$ term will rapidly oscillate and its time average will be negligible. And the Redfield operator
\begin{equation}
\begin{split}
    R_{abcd}=\frac{1}{2}[J_{acbd}(\omega_{bd})+J_{acbd}(\omega_{ac})-\\\delta_{bd}\sum_{\gamma}J_{\gamma c\gamma a}(\omega_{\gamma c})-\delta_{ac}\sum_{\gamma}J_{\gamma b\gamma d}(\omega_{\gamma d})]
\end{split}
\end{equation}
where 
\begin{equation}\label{jab}
    J_{acbd}(\omega)=\int \langle\langle a\vert A_1(0)\vert  c\rangle\langle d \vert A_1(t) \vert b\rangle\rangle e^{-i\omega t}dt
\end{equation}
which is the Fourier transform of
 the time correlation function of the spin-phonon coupling hyperfine term. 

\subsection{Second quantization form of $T_1$}
When considering the reduced spin density matrix in phonon bath, the second quantization form can be applied to the phonon displacements, then deduce the Redfield operator with a phonon Green's function automatically preserve the energy conservation, which shows as below equation\cite{doi:10.1126/sciadv.aax7163,C7SC02832F}, 
\begin{equation}\label{Redop}
\begin{split}
    R_{ab,cd}=-\frac{\pi}{2\hbar^2}\sum_v\sum_{\alpha\Vec{q}}\{\sum_j\delta_{bd}V_{aj}^{v\alpha\Vec{q}}V_{jc}^{v\alpha-\Vec{q}}G(\omega_{jc},\omega_{\alpha \Vec{q}})-\\V_{ac}^{v\alpha\Vec{q}}V_{db}^{v\alpha-\Vec{q}}G(\omega_{db},\omega_{\alpha \Vec{q}})
    -V_{ac}^{v\alpha\Vec{q}}V_{db}^{v\alpha-\Vec{q}}G(\omega_{ca},\omega_{\alpha \Vec{q}})+\\
\sum_j\delta_{ac}V_{aj}^{v\alpha\Vec{q}}V_{jb}^{v\alpha-\Vec{q}}G(\omega_{jd},\omega_{\alpha \Vec{q}})\}
    \end{split}
\end{equation}
\begin{equation}\label{greenphonon}
    G(\omega_{ij},\omega_{\alpha \Vec{q}})=\delta(\omega_{ij}-\omega_{\alpha \Vec{q}})\Bar{n}_{\alpha \Vec{q}}+\delta(\omega_{ij}+\omega_{\alpha \Vec{q}})(\Bar{n}_{\alpha \Vec{q}}+1)
\end{equation}
\begin{equation}
    V_{ab}^{v\alpha\Vec{q}}=\left\langle a \left\vert \frac{\partial H^v}{\partial Q_{\alpha\Vec{q}}}\right\vert b\right\rangle
\end{equation}

$G(\omega_{ij}, \omega_{\alpha \vec{q}})$ denotes the phonon Green's function, where $\omega_{ij}$ is the energy difference between spin states $i$ and $j$, and $\omega_{\alpha \vec{q}}$ is the phonon energy for mode $(\alpha, \vec{q})$. The delta functions contained in $G$ enforce energy conservation during phonon-mediated transitions. 

The term $\bar{n}_{\alpha \vec{q}}$ represents the average phonon occupation number for the phonon mode in branch $\alpha$ with crystal momentum $\vec{q}$, and follows the Bose–Einstein distribution function.



To calculate the spin-lattice relaxation time $T_1$ for a spin-$\frac{1}{2}$ system, it is sufficient to evaluate the dynamical equations for the population terms $\rho_{00}$ and $\rho_{11}$ of the density matrix. 

When applying the Redfield formalism, it is assumed that the timescale of the thermal bath is much shorter than the timescale of the system dynamics. This condition is satisfied in our case, allowing us to simplify the master equation by invoking the secular approximation. In this approximation, the rapidly oscillating exponential terms in Eq.~(\ref{densm}) average out to zero unless their associated energy differences vanish, effectively setting those exponentials to unity only when $\omega_{ac}+\omega_{db}=0$. 

As a result, the relaxation tensor components such as $R_{00,10}$, $R_{00,01}$, $R_{01,00}$, $R_{10,00}$, $R_{11,10}$, and $R_{11,01}$ do not contribute to the population dynamics and can be neglected in the calculation of $T_1$.

Therefore the Eq(\ref{densm}) can be simplified to 
\begin{align}\label{rho0}
    \frac{d\rho^s_{00}(t)}{dt}=R_{00,11}\rho^s_{11}(t)+R_{00,00}\rho^s_{00}(t)\nonumber
    \\
    \frac{d\rho^s_{11}(t)}{dt}=R_{11,11}\rho^s_{11}(t)+R_{11,00}\rho^s_{00}(t)
\end{align}
\begin{align}\label{r00}
  R_{00,00}=R_{11,11}=-\frac{\pi}{\hbar^2}\sum_{v\alpha\Vec{q}}V_{01}^{v\alpha\Vec{q}}V_{10}^{v\alpha-\Vec{q}}G(\omega_{10},\omega_{\alpha \Vec{q}})\nonumber
  \\
   R_{00,11}=R_{11,00}=\frac{\pi}{\hbar^2}\sum_{v\alpha\Vec{q}}V_{01}^{v\alpha\Vec{q}}V_{10}^{v\alpha-\Vec{q}}G(\omega_{10},\omega_{\alpha \Vec{q}})
\end{align}

The states $\vert 0\rangle$ and $\vert 1\rangle$ are eigenstates of the original spin Hamiltonian $H_0$. In this system, they correspond to the clock transition states 7 and 9, respectively, as reported in Ref.~\cite{Kundu2022-kg}. The eigenvalues and eigenvectors of these states were computed using the \textsc{EasySpin} package, with the spin Hamiltonian parameters listed in Table~\ref{hamilpara}.

Since we are considering only the hyperfine interaction in this analysis, the spin-phonon coupling matrix element between states $\vert 0\rangle$ and $\vert 1\rangle$ for phonon mode $(\alpha,\vec{q})$ is given by:

\begin{equation}
    V_{01}^{\alpha\vec{q}} = 
    \left\langle 0 \left\vert 
    \frac{\partial A_{xx}}{\partial Q_{\alpha\vec{q}}} \hat{S}_x \hat{I}_x +
    \frac{\partial A_{yy}}{\partial Q_{\alpha\vec{q}}} \hat{S}_y \hat{I}_y +
    \frac{\partial A_{zz}}{\partial Q_{\alpha\vec{q}}} \hat{S}_z \hat{I}_z
    \right\vert 1 \right\rangle
\end{equation}

Here, $A_{xx}$, $A_{yy}$, and $A_{zz}$ are the diagonal components of the hyperfine interaction tensor $\mathbf{A}$. The derivatives $\frac{\partial A_{ii}}{\partial Q_{\alpha\vec{q}}}$ ($i = x, y, z$) quantify the sensitivity of the hyperfine tensor to atomic displacements along the phonon mode $(\alpha, \vec{q})$, and are evaluated according to Eq.~(\ref{phodis}).


So combining the equations(\ref{rho0}),   (\ref{r00}) and Bloch equation (\ref{t1t2}), (\ref{mag}) in supplementary,
We have
\begin{equation}
    T_1=\frac{1}{R_{00,11}+R_{11,00}-R_{00,00}-R_{11,11}}=\frac{1}{4R_{00,11}}
\end{equation}

\subsection{$T_2$ due to zero-energy transfer fluctuations}
For decoherence time $T_2$, it depends on both $T_1$ relaxation and fluctuation of effective field perpendicular on the transverse plane, compare to fluctuation part, the former one is not the leading term. For $T_1$  part, with the same process of second quantization Redfield operator $\rho_{01}$ and $\rho_{10}$ should be
\begin{align}\label{rho01}
     \frac{d\rho^s_{01}(t)}{dt}=
    e^{i(\omega_{00}+\omega_{11})t}R_{01,01}\rho^s_{01}(t)\nonumber
\\
     \frac{d\rho^s_{10}(t)}{dt}=
    e^{i(\omega_{11}+\omega_{00})t}R_{10,10}\rho^s_{10}(t)
\end{align}
If follows the second quantization form, the $R_{01,01}$ shows
\begin{equation}
    R_{01,01}=-\frac{\pi}{\hbar^2}\sum_{v\alpha\Vec{q}}V_{01}^{v\alpha\Vec{q}}V_{10}^{v\alpha-\Vec{q}}G(\omega_{10},\omega_{\alpha \Vec{q}})
\end{equation}

But the energy delta function in the phonon Green's function $G(\omega_{10},\omega_{\alpha \Vec{q}})$ excludes the secular part of Redfield theory, which means $T_2$ only depends on the change of $T_1$, then the fluctuations perpendicular to the transverse plane is not taken into account.  
So with secular Redfield theory,
we have the $R_{01,01}$ as
\begin{equation}\label{r01}
\begin{split}
    R_{01,01}=J_{0011}(0)-\frac{1}{2}(J_{0000}(0)+J_{1010}(\omega)\\+J_{0101}(-\omega)+J_{1111}(0))
    \end{split}
\end{equation}
where $J_{abcd}$ is Eq.(\ref{jab}). And it is easy to derive the $R_{10,10}$ following the similar process. 

So by considering the correlation function of hyperfine tensor as
\begin{equation}\label{correlat}
    \langle A_1(0)A_1(t)\rangle=\sum_{\alpha'\Vec{q}'}\sum_{\alpha \Vec{q}} A^{\alpha'\Vec{q}'}_1 \cos(\phi_{\alpha'\Vec{q}'}) A^{\alpha\Vec{q}}_1 \cos(\omega_{\alpha\Vec{q}}t+\phi_{\alpha\Vec{q}})
\end{equation}

\(A_1\) represents the total change in the hyperfine tensor induced by all phonon modes. The quantity \(A^{\alpha\vec{q}}_1\) denotes the oscillation amplitude of the hyperfine tensor resulting from the phonon mode in branch \(\alpha\) with wave vector \(\vec{q}\), as defined in Eq.~(\ref{phodis}).  Phonon-phonon inter-mode correlations typically arise from second- or higher-order interactions. For simplicity, we consider only the time correlation function of each phonon mode with itself.
Therefore, correlations between distinct modes \((\alpha', \vec{q}')\) and \((\alpha, \vec{q})\) can be neglected. Consequently, only the diagonal terms with \(\delta_{\alpha'\vec{q}',\alpha\vec{q}}\) contribute, allowing Eq.~(\ref{correlat}) to be simplified as follows:

\begin{equation}
    \langle A_1(0)A_1(t)\rangle=\frac{1}{2}\sum_{\alpha \Vec{q}}(A^{\alpha\Vec{q}}_1)^2 \cos(\omega_{\alpha\Vec{q}}t)
\end{equation}
After incorporating the temperature dependence of phonon occupation numbers and performing the thermal ensemble average, the summation over discrete $\vec{q}$ points is converted into an integral over the Brillouin zone. The resulting phonon correlation function takes the form:

\begin{align}\label{t2cor}
 \langle\langle &A_1(0)A_1(t)\rangle\rangle\notag\\&=\frac{V}{2(2\pi)^3}\sum_{\alpha }\int dq^3 (A^{\alpha\Vec{q}}_1)^2 \cos(\omega_{\alpha\Vec{q}}t)(\frac{1}{2}+\frac{1}{e^{\hbar\omega_{\alpha\Vec{q}}/kT}-1})^2
\end{align}

V is the volume of one unit cell.
 
Since the energy of optical phonons significantly exceeds the thermal energy at room temperature, their contribution to spin-phonon relaxation can be neglected. Consequently, only the acoustic phonon branches are considered in this work. These branches exhibit approximately linear dispersion relations of the form $\omega = v \vec{q}$, where $v$ is the sound velocity.

The sound velocities of the Lu(II) complex were calculated using the \textsc{Phonopy} package, yielding values of 1854~m/s and 2275~m/s for the two transverse acoustic modes, and 2717~m/s for the longitudinal mode.

To evaluate the transverse spin relaxation time $T_2$, we combined the relevant expressions, including Eqs.~(\ref{rho01}), (\ref{r01}), (\ref{t2cor}), (\ref{t1t2}), and (\ref{mag}). The final expression for $T_2$ is given by:

\begin{equation}
    T_2 = \frac{1}{R_{01,01} - R_{10,10}}
\end{equation}

\section{Results}
\begin{figure*}[t]
\centering

\includegraphics[angle=0,width=0.9\textwidth]{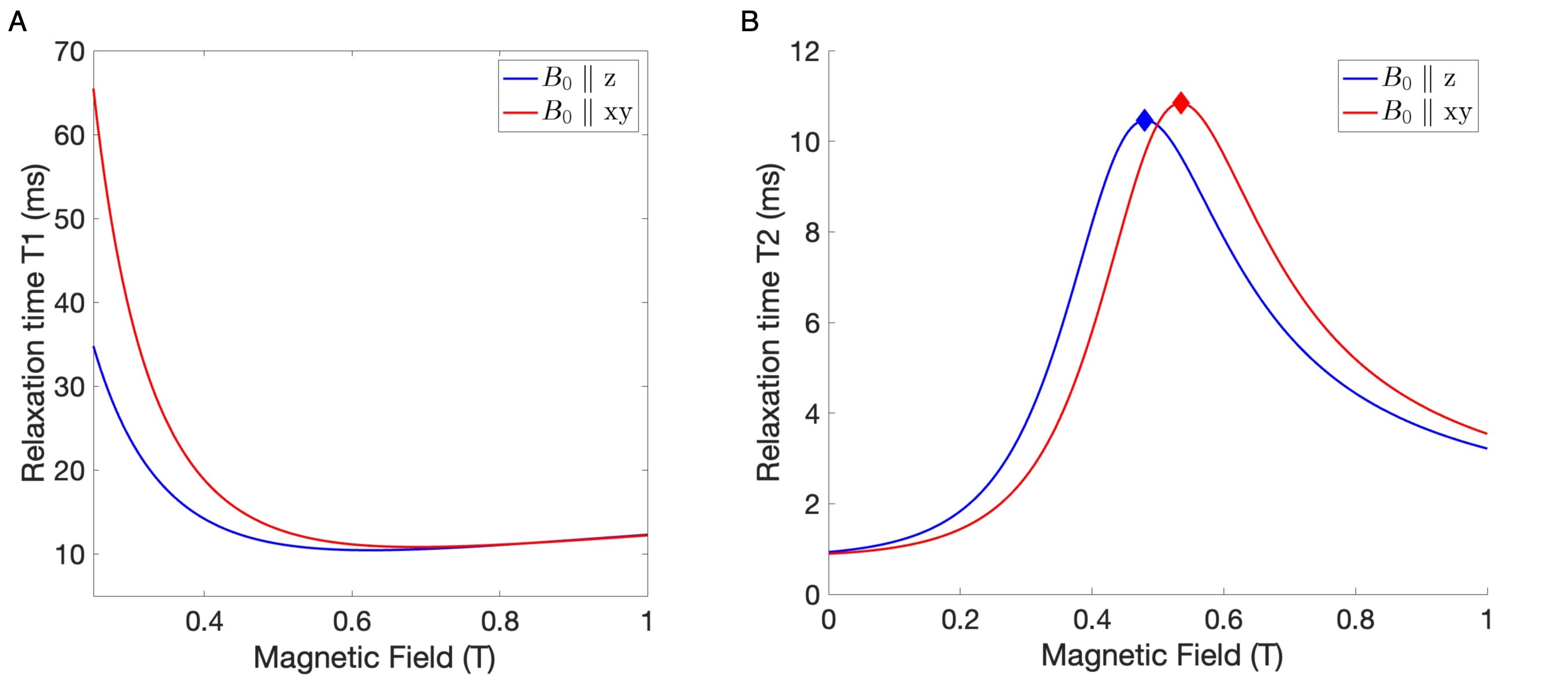}
 \caption{\textbf{A. Magnetic field dependence of longitudinal relaxation time ($T_1$) in Lu compounds.}
$T_1$ decreases sharply from zero field and reaches a minimum near 0.6 T due to changes in eigenstate composition and transition energies. \textbf{B. Variation of transverse relaxation time ($T_2$) with magnetic field for Lu compounds at 5 Kevin.} 
A peak at approximately 0.43T reflects the evolution of eigenstates and improved spin coherence under external field alignment. }
    \label{TB}
\end{figure*}
We first investigated the magnetic field dependence of both the longitudinal relaxation time ($T_1$) and the transverse relaxation time ($T_2$). By systematically scanning across different magnetic field strengths and directions, we identified a pronounced peak in $T_2$ around $B = 0.43\,\mathrm{T}$, which is in excellent agreement with the experimentally observed clock transition field. Our computed $T_2$ values not only capture this peak but also reproduce the overall experimental trend, thereby validating the accuracy of our first-principles framework in modeling spin decoherence.

The magnetic field dependence of $T_1$ and its comparison with experimental data are also shown, exhibiting consistent behavior across the field range. Following this, we evaluated the spin Hamiltonian at $B = 0.43\,\mathrm{T}$, extracting the corresponding eigenvalues and eigenstates for all subsequent analyses.

Finally, using the clock transition states at this field, we computed the temperature dependence of both $T_1$. These results offer detailed insight into the phonon-limited decoherence mechanisms in this system and provide a benchmark for the design and optimization of spin-based quantum devices.

At a temperature of 5\,K and with the magnetic field $B_0 \parallel z$, the experimentally reported spin-lattice relaxation time ($T_1$) for the clock transition lies in the range of 1.3 to 1.6\,ms. Our \textit{ab initio} calculations yield a $T_1$ value of approximately 11\,ms. For the spin coherence time ($T_2$), our result is 10.5\,ms, whereas the experimental value is approximately 12\,$\mu$s. Both $T_1$ and $T_2$ differ from experimental measurements by one to two orders of magnitude.

We also calculate the case of $B_0 \parallel xy$, both $T_1$ and $T_2$ exhibit slightly different peak positions and magnitudes compared to the $B_0 \parallel z$ configuration. This variation arises from the weak anisotropy in the Landé $g$-factor and the hyperfine tensor.


By evaluating the relaxation times \( T_1 \) and \( T_2 \) away from the clock transition point, we find that, in the absence of an external magnetic field, our calculations predict an infinite \( T_1 \). This result arises from the fact that the perturbative term \( A_1(t) \) commutes with the unperturbed spin Hamiltonian \( H_0 \), and therefore cannot induce transitions between eigenstates.

As the external magnetic field increases from zero, $T_1$ begins to decrease, reaching a minimum around 0.6 T, as Fig.~ \ref{TB}A demonstrates. This trend is attributed to the magnetic-field-driven evolution of the eigenstates, which alters both the transition matrix elements and the corresponding transition energies. Interestingly, this behavior contrasts with that of the transverse relaxation time $T_2$, which increases in the same field range, suggesting that different mechanisms dominate the two relaxation processes.

A detailed scan over the static magnetic field reveals that $T_2$ reaches a maximum at approximately $B = 0.43$ T, as shown in Fig.\ref{TB}B. This field dependence arises from two key factors. First, the eigenstates themselves evolve with the external field, modifying their projections onto the Redfield operators. Second, the energy splitting between the relevant states also changes with field strength, affecting the spectral density function $J(\omega)$ and thereby modulating the relaxation rates.




\begin{figure}[!h]
    \centering
    \includegraphics[scale=0.32]{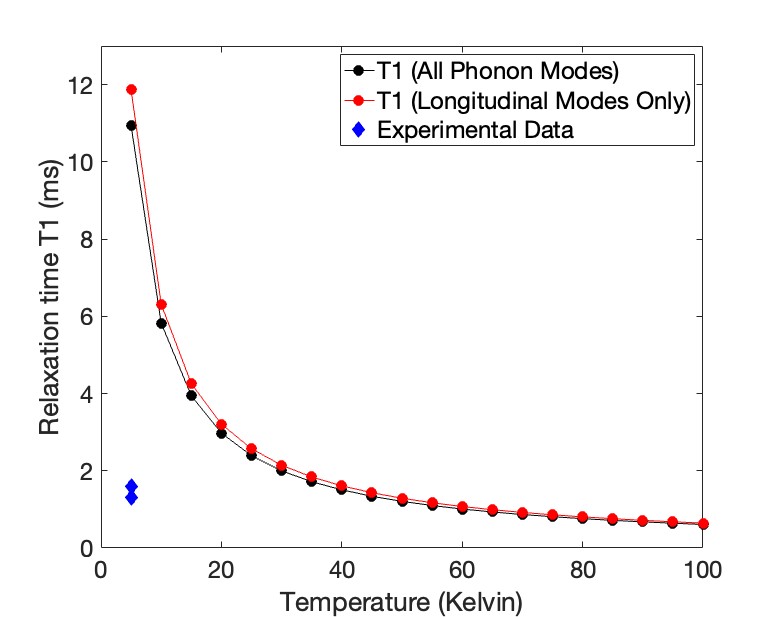}
    \caption{ Comparison between calculated $T_1$ relaxation times of Lu compounds at the clock transition ($B = 0.43$T) and experimental measurements.}
    \label{t1}
\end{figure}

The temperature dependence of \( T_1 \) is presented in Fig.~\ref{t1}, showing a gradual decrease with increasing temperature. This trend is consistent with phonon-mediated relaxation mechanisms, where the phonon occupation number follows the Bose–Einstein distribution, as described in the phonon Green's function in Eq.~\ref{greenphonon}. 

Due to energy conservation, only phonons with energies matching the transition energy between states~7 and~9 contribute significantly to the longitudinal relaxation process. Among these, numerical results indicate that longitudinal phonons dominate the contribution to \( T_1 \), even when their energies are similar to those of transverse modes. This dominance stems from the significantly larger phonon eigenvector components \( L_{is}^{\alpha\vec{q}} \) associated with longitudinal modes in Eq.~\ref{phodis}, as illustrated in Fig.~\ref{t1}.

In the case of $T_2$, our calculation shows phonon modes with intermediate wavelengths and energies within the first Brillouin zone dominate the relaxation process. The contribution from long-wavelength, low-frequency phonons is minimal because the phase difference between atoms in the unit cell is too small to substantially influence the hyperfine tensor.  Higher-energy phonons also have a limited effect due to their small magnitude and the fact that the Bose-Einstein distribution function disfavors high-energy modes at low temperatures.

\section{Discussion}

Our calculations predict that the longitudinal relaxation time \( T_1 \) diverges at zero magnetic field. However, this theoretical result contrasts with experimental findings, which consistently report finite \( T_1 \) values even at zero field. Several factors may contribute to this discrepancy.

First, while clock-transition systems are chemically engineered to localize the unpaired electron primarily in an \( s \)-orbital, the electron wavefunction inevitably includes small contributions from orbitals with non-zero angular momentum (\( l \neq 0 \)). These components enable spin-orbit coupling to mediate relaxation via phonon interactions, through mechanisms analogous to the Elliott–Yafet and Dyakonov–Perel processes~\cite{PhysRevLett.129.197201}.

Second, our theoretical framework currently employs only first-order perturbation theory. Higher-order spin-phonon interactions could introduce effective off-diagonal terms in the Hamiltonian, enabling transitions between otherwise forbidden eigenstates and leading to finite \( T_1 \) values even at zero field. Incorporating such effects in future work may help reconcile theoretical predictions with experimental observations.

Furthermore, the magnitude of our computed \( T_1 \) values differs from experimental measurements by one to two orders of magnitude. One important reason is that our model only includes phonon-induced fluctuations of the hyperfine interaction, while neglecting phonon effects on the Landé \( g \)-factor in the Zeeman term of the spin Hamiltonian [Eq.~\ref{sph}]. Prior studies~\cite{doi:10.1126/sciadv.aax7163} have shown that, particularly under strong magnetic fields, the phonon-induced modulation of \( g_e \) can dominate the relaxation process, and at intermediate fields, it contributes at a level comparable to the hyperfine-induced mechanism.

Another potential source of discrepancy lies in the relaxation theory itself. Our current analysis is based on Redfield theory, which accounts only for single-phonon (first-order) processes. However, as highlighted in prior work~\cite{doi:10.1126/sciadv.abn7880}, higher-order phonon processes such as Raman relaxation can also play a significant role in spin-lattice relaxation. These contributions are absent in the present model and may partially explain the observed mismatch with experiment.

Regarding transverse relaxation time \( T_2 \), our model includes contributions from both spin-lattice relaxation processes, represented by \( J(\omega) \), and low-frequency noise captured by the \( J(0) \) term. In this study, we considered only the time autocorrelations of individual phonon modes, assuming that inter-mode correlations are negligible.

However, in systems where phonon modes are strongly coupled or share long-time dynamical correlations, such inter-mode interactions may significantly affect decoherence. A more comprehensive theoretical treatment incorporating these correlations could further improve the accuracy of predicted \( T_2 \) values.

\section*{Acknowledgments}
This work was supported by the Center for Molecular
Magnetic Quantum Materials, an Energy Frontier Research
Center funded by the U.S. Department of Energy, Office of
Science, Basic Energy Sciences under Award No. DESC0019330.
Computations were done using the utilities
of National Energy Research Scientific Computing
Center (NERSC) and University of Florida Research Computing systems.

\onecolumngrid
\section*{Supplement}








\section{Bloch equation}
Bloch equation was developed by Felix Bloch in 1940s, which was proven to be a successful theory for nuclear magnetization relaxation\cite{PhysRev.70.460}.Over time, this formalism has been extended beyond nuclear spins to describe the dynamics of electron spins.

In modern quantum systems, especially in the context of spin-based qubits, the generalized Bloch equations provide essential insight into the time evolution of the spin density matrix under various relaxation mechanisms. These include spin-lattice relaxation ($T_1$ processes), spin dephasing ($T_2$ processes), and the influence of fluctuating environments such as phonons or other environmental bath. The Bloch formalism thus remains a foundational tool for interpreting experimental observations and validating theoretical models in quantum information science, condensed matter physics, and electron paramagnetic resonance (EPR) studies.
\begin{equation}
\frac{d\mathbf{M}}{dt} = \gamma \mathbf{M} \times \mathbf{B} - \frac{M_x}{T_2}\hat{i} - \frac{M_y}{T_2}\hat{j} - \frac{M_z - M_0}{T_1}\hat{k}
\end{equation}

In component form, the equations are:

\begin{align}\label{t1t2}
\frac{dM_x}{dt} &= \gamma (M_y B_z - M_z B_y) - \frac{M_x}{T_2} \nonumber\\
\frac{dM_y}{dt} &= \gamma (M_z B_x - M_x B_z) - \frac{M_y}{T_2} \nonumber\\
\frac{dM_z}{dt} &= \gamma (M_x B_y - M_y B_x) - \frac{M_z - M_0}{T_1}
\end{align}
$\gamma$ is the electron gyromagnetic ratio. 
The magnetic moment in the spin density form is 
\begin{align}\label{mag}
    M_z=\rho_{00}-\rho_{11}\nonumber\\
    M_x=\frac{1}{2}(\rho_{01}+\rho_{10})\nonumber\\
     M_y=\frac{i}{2}(\rho_{01}-\rho_{10})
\end{align}
\section{DFT calculation and hyperfine tensor variation fitting }

To evaluate Eq.~(\ref{phodis}), each atom in the unit cell was displaced incrementally from $-0.008$ to $0.008$ Å along the three crystallographic directions ($a$, $b$, and $c$), with six displacement steps per direction. Each step size was approximately 2.67$\times 10^{-3}$~ Å. For each displaced configuration, DFT calculations were performed to obtain the corresponding hyperfine tensor.

\begin{figure}
    \centering
    \includegraphics[width=1\linewidth]{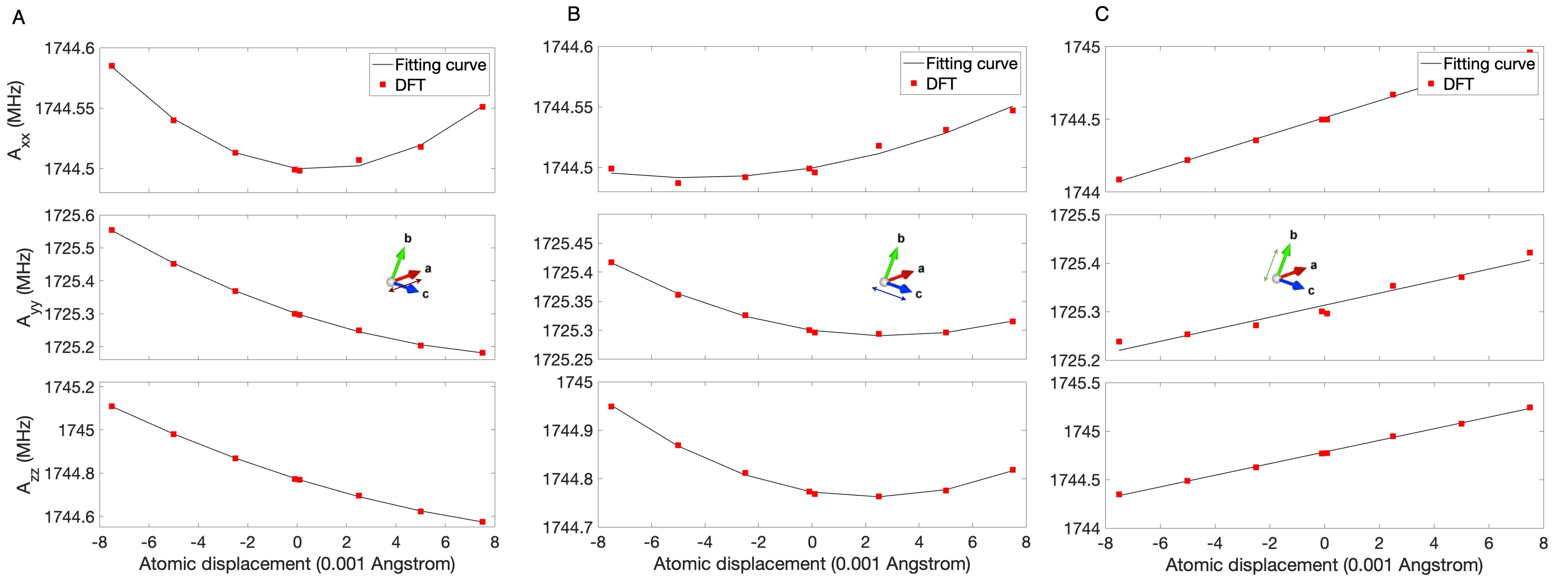}
    \caption{\textbf{Hyperfine tensor response to atomic displacements along crystallographic directions.} 
\textbf{A.} Displacement of the Lu atom along the lattice vector \textbf{a} induces changes in the hyperfine tensor components. 
\textbf{B.} Displacement along the lattice vector \textbf{c} produces corresponding variations. 
\textbf{C.} Displacement along the lattice vector \textbf{b} results in further tensor modifications.}
    \label{displace}
\end{figure}

The discrete variations in the hyperfine tensor components resulting from atomic displacements were fitted with second-order polynomial functions for each atom and displacement direction, allowing smooth interpolation of the hyperfine tensor as a function of atomic motion. To minimize numerical artifacts, the second-order term was excluded if the phonon-induced displacement of a given atom exceeded 50\% of the maximum displacement range used in the original DFT calculations. An example of the fitted curves for one Lu atom is shown in Fig.~\ref{displace}.

\bibliography{spinrelax}

\begin{thebibliography}{10}
\providecommand{\url}[1]{#1}
\csname url@samestyle\endcsname
\providecommand{\newblock}{\relax}
\providecommand{\bibinfo}[2]{#2}
\providecommand{\BIBentrySTDinterwordspacing}{\spaceskip=0pt\relax}
\providecommand{\BIBentryALTinterwordstretchfactor}{4}
\providecommand{\BIBentryALTinterwordspacing}{\spaceskip=\fontdimen2\font plus
\BIBentryALTinterwordstretchfactor\fontdimen3\font minus
  \fontdimen4\font\relax}
\providecommand{\BIBforeignlanguage}[2]{{%
\expandafter\ifx\csname l@#1\endcsname\relax
\typeout{** WARNING: IEEEtran.bst: No hyphenation pattern has been}%
\typeout{** loaded for the language `#1'. Using the pattern for}%
\typeout{** the default language instead.}%
\else
\language=\csname l@#1\endcsname
\fi
#2}}
\providecommand{\BIBdecl}{\relax}
\BIBdecl

\bibitem{PhysRevB.65.205309}
\BIBentryALTinterwordspacing
I.~A. Merkulov, A.~L. Efros, and M.~Rosen, ``Electron spin relaxation by nuclei
  in semiconductor quantum dots,'' \emph{Phys. Rev. B}, vol.~65, p. 205309, Apr
  2002. [Online]. Available:
  \url{https://link.aps.org/doi/10.1103/PhysRevB.65.205309}
\BIBentrySTDinterwordspacing

\bibitem{Saykin2002}
\BIBentryALTinterwordspacing
S.~Saykin, D.~Mozyrsky, and V.~Privman, ``Relaxation of shallow donor electron
  spin due to interaction with nuclear spin bath,'' \emph{Nano Letters},
  vol.~2, no.~6, pp. 651--655, Jun 2002. [Online]. Available:
  \url{https://doi.org/10.1021/nl0255552}
\BIBentrySTDinterwordspacing

\bibitem{PhysRevB.74.195301}
\BIBentryALTinterwordspacing
W.~Yao, R.-B. Liu, and L.~J. Sham, ``Theory of electron spin decoherence by
  interacting nuclear spins in a quantum dot,'' \emph{Phys. Rev. B}, vol.~74,
  p. 195301, Nov 2006. [Online]. Available:
  \url{https://link.aps.org/doi/10.1103/PhysRevB.74.195301}
\BIBentrySTDinterwordspacing

\bibitem{PhysRevB.85.115303}
\BIBentryALTinterwordspacing
N.~Zhao, S.-W. Ho, and R.-B. Liu, ``Decoherence and dynamical decoupling
  control of nitrogen vacancy center electron spins in nuclear spin baths,''
  \emph{Phys. Rev. B}, vol.~85, p. 115303, Mar 2012. [Online]. Available:
  \url{https://link.aps.org/doi/10.1103/PhysRevB.85.115303}
\BIBentrySTDinterwordspacing

\bibitem{PhysRevLett.110.160402}
\BIBentryALTinterwordspacing
L.~Ratschbacher, C.~Sias, L.~Carcagni, J.~M. Silver, C.~Zipkes, and M.~K\"ohl,
  ``Decoherence of a single-ion qubit immersed in a spin-polarized atomic
  bath,'' \emph{Phys. Rev. Lett.}, vol. 110, p. 160402, Apr 2013. [Online].
  Available: \url{https://link.aps.org/doi/10.1103/PhysRevLett.110.160402}
\BIBentrySTDinterwordspacing

\bibitem{doi:10.1021/jacs.1c05068}
\BIBentryALTinterwordspacing
M.~Briganti, F.~Santanni, L.~Tesi, F.~Totti, R.~Sessoli, and A.~Lunghi, ``A
  complete ab initio view of orbach and raman spin–lattice relaxation in a
  dysprosium coordination compound,'' \emph{Journal of the American Chemical
  Society}, vol. 143, no.~34, pp. 13\,633--13\,645, 2021, pMID: 34465096.
  [Online]. Available: \url{https://doi.org/10.1021/jacs.1c05068}
\BIBentrySTDinterwordspacing

\bibitem{Murali2003APF}
\BIBentryALTinterwordspacing
N.~Murali and V.~V. Krishnan, ``A primer for nuclear magnetic relaxation in
  liquids,'' \emph{Concepts in Magnetic Resonance Part A}, vol.~17, pp.
  86--116, 2003. [Online]. Available:
  \url{https://api.semanticscholar.org/CorpusID:36481129}
\BIBentrySTDinterwordspacing

\bibitem{STAVROU2021114605}
\BIBentryALTinterwordspacing
V.~Stavrou, ``Phonon-induced decoherence of a spin based qubit made with
  asymmetric coupled quantum dots,'' \emph{Physica E: Low-dimensional Systems
  and Nanostructures}, vol. 130, p. 114605, 2021. [Online]. Available:
  \url{https://www.sciencedirect.com/science/article/pii/S1386947720316738}
\BIBentrySTDinterwordspacing

\bibitem{PhysRevB.101.165302}
\BIBentryALTinterwordspacing
P.~Stipsi\ifmmode~\acute{c}\else \'{c}\fi{} and
  M.~Milivojevi\ifmmode~\acute{c}\else \'{c}\fi{}, ``Control of a spin qubit in
  a lateral gaas quantum dot based on symmetry of gating potential,''
  \emph{Phys. Rev. B}, vol. 101, p. 165302, Apr 2020. [Online]. Available:
  \url{https://link.aps.org/doi/10.1103/PhysRevB.101.165302}
\BIBentrySTDinterwordspacing

\bibitem{PhysRevA.100.042309}
\BIBentryALTinterwordspacing
F.~M. Souza, P.~A. Oliveira, and L.~Sanz, ``Quantum entanglement driven by
  electron-vibrational mode coupling,'' \emph{Phys. Rev. A}, vol. 100, p.
  042309, Oct 2019. [Online]. Available:
  \url{https://link.aps.org/doi/10.1103/PhysRevA.100.042309}
\BIBentrySTDinterwordspacing

\bibitem{PhysRevB.98.075403}
\BIBentryALTinterwordspacing
M.~Gawe\l{}czyk, M.~Krzykowski, K.~Gawarecki, and P.~Machnikowski,
  ``Controllable electron spin dephasing due to phonon state distinguishability
  in a coupled quantum dot system,'' \emph{Phys. Rev. B}, vol.~98, p. 075403,
  Aug 2018. [Online]. Available:
  \url{https://link.aps.org/doi/10.1103/PhysRevB.98.075403}
\BIBentrySTDinterwordspacing

\bibitem{PhysRevLett.93.016601}
\BIBentryALTinterwordspacing
V.~N. Golovach, A.~Khaetskii, and D.~Loss, ``Phonon-induced decay of the
  electron spin in quantum dots,'' \emph{Phys. Rev. Lett.}, vol.~93, p. 016601,
  Jun 2004. [Online]. Available:
  \url{https://link.aps.org/doi/10.1103/PhysRevLett.93.016601}
\BIBentrySTDinterwordspacing

\bibitem{Gomez-Coca2014}
\BIBentryALTinterwordspacing
S.~G{\'o}mez-Coca, A.~Urtizberea, E.~Cremades, P.~J. Alonso, A.~Cam{\'o}n,
  E.~Ruiz, and F.~Luis, ``Origin of slow magnetic relaxation in kramers ions
  with non-uniaxial anisotropy,'' \emph{Nature Communications}, vol.~5, no.~1,
  p. 4300, Jul 2014. [Online]. Available:
  \url{https://doi.org/10.1038/ncomms5300}
\BIBentrySTDinterwordspacing

\bibitem{PhysRevB.101.045202}
\BIBentryALTinterwordspacing
J.~Park, J.-J. Zhou, and M.~Bernardi, ``Spin-phonon relaxation times in
  centrosymmetric materials from first principles,'' \emph{Phys. Rev. B}, vol.
  101, p. 045202, Jan 2020. [Online]. Available:
  \url{https://link.aps.org/doi/10.1103/PhysRevB.101.045202}
\BIBentrySTDinterwordspacing

\bibitem{PhysRevLett.129.197201}
\BIBentryALTinterwordspacing
J.~Park, J.-J. Zhou, Y.~Luo, and M.~Bernardi, ``Predicting phonon-induced spin
  decoherence from first principles: Colossal spin renormalization in condensed
  matter,'' \emph{Phys. Rev. Lett.}, vol. 129, p. 197201, Nov 2022. [Online].
  Available: \url{https://link.aps.org/doi/10.1103/PhysRevLett.129.197201}
\BIBentrySTDinterwordspacing

\bibitem{PhysRevB.106.174404}
\BIBentryALTinterwordspacing
J.~Park, Y.~Luo, J.-J. Zhou, and M.~Bernardi, ``Many-body theory of
  phonon-induced spin relaxation and decoherence,'' \emph{Phys. Rev. B}, vol.
  106, p. 174404, Nov 2022. [Online]. Available:
  \url{https://link.aps.org/doi/10.1103/PhysRevB.106.174404}
\BIBentrySTDinterwordspacing

\bibitem{Lunghi2017}
\BIBentryALTinterwordspacing
A.~Lunghi, F.~Totti, R.~Sessoli, and S.~Sanvito, ``The role of anharmonic
  phonons in under-barrier spin relaxation of single molecule magnets,''
  \emph{Nature Communications}, vol.~8, no.~1, p. 14620, Mar 2017. [Online].
  Available: \url{https://doi.org/10.1038/ncomms14620}
\BIBentrySTDinterwordspacing

\bibitem{C7SC02832F}
\BIBentryALTinterwordspacing
A.~Lunghi, F.~Totti, S.~Sanvito, and R.~Sessoli, ``Intra-molecular origin of
  the spin-phonon coupling in slow-relaxing molecular magnets,'' \emph{Chem.
  Sci.}, vol.~8, pp. 6051--6059, 2017. [Online]. Available:
  \url{http://dx.doi.org/10.1039/C7SC02832F}
\BIBentrySTDinterwordspacing

\bibitem{Escalera-Moreno2017}
\BIBentryALTinterwordspacing
L.~Escalera-Moreno, N.~Suaud, A.~Gaita-Ari{\~{n}}o, and E.~Coronado,
  ``Determining key local vibrations in the relaxation of molecular spin qubits
  and single-molecule magnets,'' \emph{The Journal of Physical Chemistry
  Letters}, vol.~8, no.~7, pp. 1695--1700, Apr 2017. [Online]. Available:
  \url{https://doi.org/10.1021/acs.jpclett.7b00479}
\BIBentrySTDinterwordspacing

\bibitem{doi:10.1126/sciadv.aax7163}
\BIBentryALTinterwordspacing
A.~Lunghi and S.~Sanvito, ``How do phonons relax molecular spins?''
  \emph{Science Advances}, vol.~5, no.~9, p. eaax7163, 2019. [Online].
  Available: \url{https://www.science.org/doi/abs/10.1126/sciadv.aax7163}
\BIBentrySTDinterwordspacing

\bibitem{Briganti2021}
\BIBentryALTinterwordspacing
M.~Briganti, F.~Santanni, L.~Tesi, F.~Totti, R.~Sessoli, and A.~Lunghi, ``A
  complete ab initio view of orbach and raman spin--lattice relaxation in a
  dysprosium coordination compound,'' \emph{Journal of the American Chemical
  Society}, vol. 143, no.~34, pp. 13\,633--13\,645, Sep 2021. [Online].
  Available: \url{https://doi.org/10.1021/jacs.1c05068}
\BIBentrySTDinterwordspacing

\bibitem{doi:10.1126/sciadv.abn7880}
\BIBentryALTinterwordspacing
A.~Lunghi, ``Toward exact predictions of spin-phonon relaxation times: An ab
  initio implementation of open quantum systems theory,'' \emph{Science
  Advances}, vol.~8, no.~31, p. eabn7880, 2022. [Online]. Available:
  \url{https://www.science.org/doi/abs/10.1126/sciadv.abn7880}
\BIBentrySTDinterwordspacing

\bibitem{Shiddiq2016}
\BIBentryALTinterwordspacing
M.~Shiddiq, D.~Komijani, Y.~Duan, A.~Gaita-Ari{\~{n}}o, E.~Coronado, and
  S.~Hill, ``Enhancing coherence in molecular spin qubits via atomic clock
  transitions,'' \emph{Nature}, vol. 531, no. 7594, pp. 348--351, Mar 2016.
  [Online]. Available: \url{https://doi.org/10.1038/nature16984}
\BIBentrySTDinterwordspacing

\bibitem{Kundu2022-kg}
K.~Kundu, J.~R.~K. White, S.~A. Moehring, J.~M. Yu, J.~W. Ziller, F.~Furche,
  W.~J. Evans, and S.~Hill, ``A {9.2-GHz} clock transition in a {Lu(II})
  molecular spin qubit arising from a {3,467-MHz} hyperfine interaction,''
  \emph{Nature Chemistry}, vol.~14, no.~4, pp. 392--397, Apr. 2022.

\bibitem{Gakiya-Teruya2025}
\BIBentryALTinterwordspacing
M.~Gakiya-Teruya, R.~Stewart, L.~Peng, S.~Liu, C.~Li, H.-P. Cheng, G.~K.-L.
  Chan, S.~Hill, and M.~Shatruk, ``A 54.6 ghz clock transition in ho3+ electron
  spin qubits assembled into a metal--organic framework,'' \emph{Journal of the
  American Chemical Society}, Jun 2025. [Online]. Available:
  \url{https://doi.org/10.1021/jacs.5c07796}
\BIBentrySTDinterwordspacing

\bibitem{PhysRevB.73.155311}
\BIBentryALTinterwordspacing
M.~Borhani, V.~N. Golovach, and D.~Loss, ``Spin decay in a quantum dot coupled
  to a quantum point contact,'' \emph{Phys. Rev. B}, vol.~73, p. 155311, Apr
  2006. [Online]. Available:
  \url{https://link.aps.org/doi/10.1103/PhysRevB.73.155311}
\BIBentrySTDinterwordspacing

\bibitem{Kornich2018phononassisted}
\BIBentryALTinterwordspacing
V.~Kornich, C.~Kloeffel, and D.~Loss, ``Phonon-assisted relaxation and
  decoherence of singlet-triplet qubits in {S}i/{S}i{G}e quantum dots,''
  \emph{{Quantum}}, vol.~2, p.~70, May 2018. [Online]. Available:
  \url{https://doi.org/10.22331/q-2018-05-28-70}
\BIBentrySTDinterwordspacing

\bibitem{YU2022111000}
\BIBentryALTinterwordspacing
Y.~Yu, X.~Zhang, S.~Dillon, J.~Chen, Y.~Chen, H.-P. Cheng, and X.-G. Zhang,
  ``Ampere field fluctuation from acoustic phonons as a possible source of spin
  decoherence,'' \emph{Journal of Physics and Chemistry of Solids}, vol. 171,
  p. 111000, 2022. [Online]. Available:
  \url{https://www.sciencedirect.com/science/article/pii/S0022369722004176}
\BIBentrySTDinterwordspacing

\bibitem{STOLL200642}
\BIBentryALTinterwordspacing
S.~Stoll and A.~Schweiger, ``Easyspin, a comprehensive software package for
  spectral simulation and analysis in epr,'' \emph{Journal of Magnetic
  Resonance}, vol. 178, no.~1, pp. 42--55, 2006. [Online]. Available:
  \url{https://www.sciencedirect.com/science/article/pii/S1090780705002892}
\BIBentrySTDinterwordspacing

\bibitem{89596ae1770f47d59ee96c48b2b10c8f}
S.~Stoll and D.~Goldfarb, ``\BIBforeignlanguage{English}{Epr interactions -
  nuclear quadrupole couplings},''
  \emph{\BIBforeignlanguage{English}{eMagRes}}, vol.~6, no.~4, pp. 495--510,
  Dec. 2017.

\bibitem{phonopy-phono3py-JPCM}
A.~Togo, L.~Chaput, T.~Tadano, and I.~Tanaka, ``Implementation strategies in
  phonopy and phono3py,'' \emph{J. Phys. Condens. Matter}, vol.~35, no.~35, p.
  353001, 2023.

\bibitem{PhysRevB.49.14251}
\BIBentryALTinterwordspacing
G.~Kresse and J.~Hafner, ``Ab initio molecular-dynamics simulation of the
  liquid-metal--amorphous-semiconductor transition in germanium,'' \emph{Phys.
  Rev. B}, vol.~49, pp. 14\,251--14\,269, May 1994. [Online]. Available:
  \url{https://link.aps.org/doi/10.1103/PhysRevB.49.14251}
\BIBentrySTDinterwordspacing

\bibitem{PhysRevB.50.17953}
\BIBentryALTinterwordspacing
P.~E. Bl\"ochl, ``Projector augmented-wave method,'' \emph{Phys. Rev. B},
  vol.~50, pp. 17\,953--17\,979, Dec 1994. [Online]. Available:
  \url{https://link.aps.org/doi/10.1103/PhysRevB.50.17953}
\BIBentrySTDinterwordspacing

\bibitem{PhysRevB.44.943}
\BIBentryALTinterwordspacing
V.~I. Anisimov, J.~Zaanen, and O.~K. Andersen, ``Band theory and mott
  insulators: Hubbard u instead of stoner i,'' \emph{Phys. Rev. B}, vol.~44,
  pp. 943--954, Jul 1991. [Online]. Available:
  \url{https://link.aps.org/doi/10.1103/PhysRevB.44.943}
\BIBentrySTDinterwordspacing

\bibitem{Rohrbach_2003}
\BIBentryALTinterwordspacing
A.~Rohrbach, J.~Hafner, and G.~Kresse, ``Electronic correlation effects in
  transition-metal sulfides,'' \emph{Journal of Physics: Condensed Matter},
  vol.~15, no.~6, p. 979, feb 2003. [Online]. Available:
  \url{https://dx.doi.org/10.1088/0953-8984/15/6/325}
\BIBentrySTDinterwordspacing

\bibitem{PhysRevB.94.014104}
\BIBentryALTinterwordspacing
E.~L. da~Silva, A.~G. Marinopoulos, R.~B.~L. Vieira, R.~C. Vil\~ao, H.~V.
  Alberto, J.~M. Gil, R.~L. Lichti, P.~W. Mengyan, and B.~B. Baker,
  ``Electronic structure of interstitial hydrogen in lutetium oxide from
  $\text{DFT}+u$ calculations and comparison study with
  $\ensuremath{\mu}\mathrm{SR}$ spectroscopy,'' \emph{Phys. Rev. B}, vol.~94,
  p. 014104, Jul 2016. [Online]. Available:
  \url{https://link.aps.org/doi/10.1103/PhysRevB.94.014104}
\BIBentrySTDinterwordspacing

\bibitem{PhysRev.70.460}
\BIBentryALTinterwordspacing
F.~Bloch, ``Nuclear induction,'' \emph{Phys. Rev.}, vol.~70, pp. 460--474, Oct
  1946. [Online]. Available:
  \url{https://link.aps.org/doi/10.1103/PhysRev.70.460}
\BIBentrySTDinterwordspacing

\end{thebibliography}
\end{document}